\documentclass[aps,prb,twocolumn,amsfonts,showpacs,showkeys,superscriptaddress,byrevtex,nofootinbib,citeautoscript]{revtex4-1}
\usepackage[dvips]{graphicx}
\usepackage{natbib}

\newcommand*\mapbi{CH$_3$NH$_3$PbI$_3$}
\newcommand*\MA{CH$_3$NH$_3^+$}

\begin{document}
\title{Non-hydrogenic excitons in perovskite CH$_3$NH$_3$PbI$_3$}
\author{E. \surname{Men\'{e}ndez-Proupin}}
\affiliation{Departamento de F\'isica, Facultad de Ciencias, 
Universidad de Chile, Casilla 653, Santiago, Chile}
\affiliation{Instituto de Energ\'ia Solar and Dept. TFB, E.T.S.I. Telecomunicaci\'on, Universidad Polit\'ecnica de Madrid, Spain}
\author{Carlos L. \surname{Beltr\'an R\'ios}}
\affiliation{Escuela de F\'isica, Universidad Industrial de Santander, A. A. 678, Bucaramanga, Colombia}
\author{P. \surname{Wahn\'on}}
\affiliation{Instituto de Energ\'ia Solar and Dept. TFB, E.T.S.I. Telecomunicaci\'on, Universidad Polit\'ecnica de Madrid, Spain}

\date{\today}

\begin{abstract}
The excitons in the orthorhombic phase of the perovskite  CH$_3$NH$_3$PbI$_3$  are studied using the effective mass approximation. The electron-hole interaction is screened by  a distance-dependent dielectric function, as described by the Haken potential or the Pollmann-B\"uttner potential. The energy spectrum and the eigenfunctions are calculated for both cases. The effective masses, the low and high frequency dielectric constants, and the interband absorption matrix elements, are obtained from generalized density functional theory calculations. 
The results show that the Pollmann-B\"uttner model provides better agreement with the experimental results. 
The discrete part of the exciton spectrum is composed of a fundamental state with a binding energy of 
24 meV, and higher states that are within 2 meV from the onset the unbound exciton continuum. 
Light absorption is dominated by the fundamental line with an oscillator strength of 0.013, followed by the exciton continuum. 
The calculations have been performed without fitting any parameter from experiments and are in close agreement with recent experimental results. 
\end{abstract}
\maketitle


Hybrid organic/inorganic perovskites based on lead or tin tri-halides are semiconductor materials that have revolutionized the research of thin film solar 
cells. With the first prototypes demonstrated six years ago\cite{kojima2009}, record cell  
efficiencies have surpassed the barrier of 20\%\cite{nrelchart,Jacoby2015}.  
Methyl-ammonium lead iodide (\mapbi{}) is one of the most studied members of this family,
 and it has been applied as photon absorber and charge transporting material\cite{ja307789s,jz400892a}.  
 
\mapbi{} presents two phase transitions at $\sim 162$~K and $\sim 327$~K. At these transitions, 
the crystal symmetry changes first from orthorhombic to tetragonal and then to cubic symmetry\cite{knop89part2,baikie2013,stoumpos2013,kawamura2002}. The three phases differ by small changes 
of the lattice vectors, rotations of the characteristic PbI$_6$ octahedra, and the orientation of the \MA{} 
cations. In the tetragonal and cubic phases, the \MA{} cations present orientational and dynamic disorder\cite{wasylishen85}, with a deep effect on the  dielectric properties\cite{lin2015,frost2014}.
In the low temperature orthorhombic phase, the \MA{} positions and orientations are fixed\cite{knop89part2,baikie2013}. 

The electronic band structure of \mapbi{} has been explained on the basis 
of generalized density functional theory (hybrid functionals)  
or Green functions GW calculations, in both cases including the spin-orbit 
coupling\cite{umarimapbi14,brivio2014,mapbi3_1}. 
For the orthorhombic phase, the valence band maximum (VBM) and the conduction band minimum (CBM) 
are located at the $\Gamma$ point corresponding to the 48-atoms unit cell, and the 
fundamental gap is 1.68~eV\cite{ishihara94}. Both the VBM and CBM are doubly degenerated, with 
nearly symmetric effective mass tensors. 

Exciton peaks are observed in the light absorption spectra at low
 temperature\cite{hirasawa94a,Tanaka2003,bindingmapbi}, just below the interband absorption edge, or melded with it, 
 depending on the temperature.   
 According to the Wannier-Mott model\cite{wannier,knox}, the exciton is similar to a hydrogen atom with the 
 proton and electron masses replaced by the hole and electron effective masses, and the Coulomb interaction 
 is screened by a dielectric constant $\epsilon$. Therefore, the exciton binding energy and the Bohr radius  are
  $Ry=\mu e^4/2\hbar^2\epsilon^{2}$ and $a_{ex}=\hbar^2\epsilon/\mu e^2$, where $\mu=m_e m_h/(m_e+m_h)$ is 
  the reduced electron-hole mass. 

One distinct feature of \mapbi{}  is the large difference between the static dielectric constant $\epsilon_{0}$
and the high frequency constant $\epsilon_{\infty}$, i.e., for frequencies higher than those of the phonon absorption. 
Values of  $\epsilon_{\infty}$  in the range $4.5-6.5$ have been calculated\cite{walsh2013,umarimapbi14,mapbi3_1,brivio2014}, while values close to 25 have been   
estimated for $\epsilon_0$\cite{walsh2013,brivio2014}.
Such difference is larger than in traditional inorganic semiconductor and should cause important polaron 
effects, such as the effective mass and gap renormalization, as well as and non-hydrogenic exciton states.  
For the latter, immediately arises the question wether  the screening constant 
$\epsilon$ should be the static dielectric constant $\epsilon_0$ or the high frequency 
$\epsilon_{\infty}$. Using the values listed in Table \ref{tab2}, the static and the high frequency dielectric constants lead to very different values of the exciton binding energy $Ry_0=2.8$~meV and $Ry_{\infty}=50$~meV, respectively. Such different energies lead to different conclusions with respect to exciton dissociation due to thermal excitation, as well as 
to different interpretation of luminescence and transport properties. 

Early estimations of the exciton binding energy\cite{hirasawa94a,Tanaka2003} $\sim 37-50$~meV were based on measurement of the exciton diamagnetic coefficient and interpretation based on the hydrogenic model with screening by 
 $\epsilon_{\infty}=6.5$.
Recent studies of the temperature dependence of photoluminescence spectra\cite{Sun2014perovskitas}, 
and numerical analysis of the absorption spectra\cite{evenjpcl14,portugall2015}  have 
provided updated  exciton binding energies around 16-19~meV. The latter values point to a screening 
constant  intermediate between $\epsilon_{\infty}$ and $\epsilon_{0}$. 
Even et al\cite{evenjpcl14} fitted the absorption spectrum using the 
Wannier-Mott exciton model
and obtained an effective dielectric constant $\epsilon_{eff}=11$.

In fact, the differences between $\epsilon_0$ and 
$\epsilon_{\infty}$ express  the electric polarization associated to the optical phonons and the electron-phonon interaction. The stationary states are coupled states of electronic and the vibrational phonon field. 
The quantum calculation of these coupled states is beyond the current capabilities of ab initio methods. 
Model Hamiltonians\cite{haken1956,haken1958,haken-pollmann-buttner} 
allow one to map the coupled electron-phonon excitations into effective electronic states, and to 
obtain the energies of stationary states.  Even when simplifying approximations are inherent in the 
models, they can provide a criterium on the relevant dielectric screening constants.  
In this Article, we apply the model Hamiltonians of Haken\cite{haken1956,haken1958} and that of 
 Pollmann and B\"uttner\cite{haken-pollmann-buttner} to the exciton spectrum. 
This formalism is applicable to the low temperature orthorhombic phase because 
in the tetragonal phase the 
static dielectric increases strongly, associated to the reorientation of \MA{} cations, and the exciton effects
practically disappear\cite{lin2015,frost2014,evenjpcl14,portugall2015}. 

 
The strength of the interaction of electrons and optical phonons is 
given by the  coupling constant 
\begin{equation}
\label{ec:coupling}
\alpha_p=\sqrt{m e^4/2\hbar^2 \epsilon_{*}^2 E_{LO}},
\end{equation}
 where $E_{LO}$ is the energy of the longitudinal optical phonon. This model was developed
 for simple crystals that display one single LO phonon branch. For this application, we have chosen 
 $E_{LO}$ as the shift of the main peak in the \mapbi{} Raman spectrum\cite{ramanmapbi3}. 
The ionic screening parameter appearing in Eq. (\ref{ec:coupling}) is 
$1/\epsilon_{*}=1/\epsilon_{\infty}-1/\epsilon_{0}$. 

For transport properties, relevant after exciton dissociation, polaron masses must be considered rather  
 than the bare electronic masses computed with fixed ions. 
They can be estimated using the  Fr\"ohlich's continuum theory of the large 
polaron\cite{polaronreview}, which predicts  
$$m^{*}=m\left( 1+ \frac{\alpha_p}{6} \right). $$
The polaron bands undergo an additional shift given by $\Delta E_p=-\alpha_p E_{LO}$. With the 
data of Table \ref{tab2}, this leads to a reduction of the electronic band gap by 95 meV.

\begin{figure}[tt!]
\begin{center}
\includegraphics[width=8.0cm]{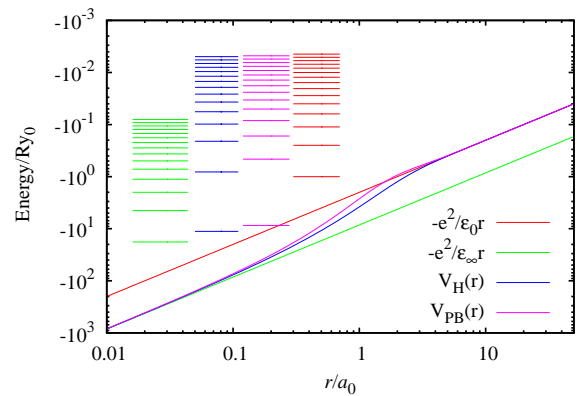}
\caption{Haken and PB potentials compared with the Coulomb potential screened by
 $\epsilon_0$ and $\epsilon_{\infty}$. Also shown are the eigenenergies for each potential.\label{fig:plotpot}}
\end{center}
\end{figure}

The Haken model\cite{haken1956,haken1958} describes two interacting polarons, each one with a radius much smaller than the exciton effective radius, and expresses the effective potential for the electron-hole Coulomb interaction as
\begin{equation}
V_H(r)=-\frac{e^2}{\epsilon_0 r} -\frac{e^2}{2\epsilon_{*} r}\left( e^{-r/l_e}  + e^{-r/l_h} \right) .
\end{equation}
Here $l_{e,h}=\sqrt{\hbar^2/2m_{e,h}E_{lo}}$ ￼are the electron- and hole-polaron radii determined using ‘‘bare’’ band electron and hole effective masses. The polaron effective mass parameters must be used in the kinetic energy terms of the Hamiltonian\cite{hakenfrances}. 

The model proposed by Pollmann and B\"uttner\cite{haken-pollmann-buttner} (PB) takes into
account the correlation between electron and hole polarons, and leads to corrections to the Haken potential. 
The resulting electron-hole interaction potential is
\begin{equation}
V_{PB}(r)=-\frac{e^2}{\epsilon_0 r} 
  -\frac{e^2}{\epsilon_{*} r}\left( \frac{m_h}{\Delta m} e^{-r/l_h}  - \frac{m_e}{\Delta m} e^{-r/l_e}\right),
\end{equation}
with $\Delta m=m_h-m_e$. This potential was derived assuming $l_{e,h}$ that the polaron lengths are 
much smaller than the effective exciton radius, which entered as a variational parameter in the original 
calculations\cite{haken-pollmann-buttner}.
The \textit{bare} band electron and hole masses must be used in the kinetic energy terms of the PB Hamiltonian.

\begin{table}
\caption{ \label{tab2}
Parameters defining the polarons in \mapbi.}
\begin{ruledtabular}
\begin{tabular}{lrrr}
 \multicolumn{4}{l}{Dielectric constants} \\
$\epsilon_{\infty}$\ & 5.32 \footnotemark[1]   \\
$\epsilon_{\infty}/\epsilon_0$ & 0.236  \footnotemark[2] \\
 \multicolumn{4}{l}{LO phonon energy} \\
$E_{LO}$ & 38.5 meV \footnotemark[3] \\
 \multicolumn{4}{l}{Coupling constants} \\
$\alpha_e$ & 1.18   \\
$ \alpha_h$ &  1.28   \\
 \multicolumn{4}{l}{Bare carrier masses} \\
$m_e/m_0$\footnotemark[1] &  0.190   \\
$ m_h/m_0$\footnotemark[1] &  0.225    \\
 \multicolumn{4}{l}{Polaron  masses} \\
$m_e^{*}/m_0$&  0.228  \\
$ m_h^{*}/m_0$ &  0.273    \\
 \multicolumn{4}{l}{Polaron  radii} \\
$ l_e$&  22.83 \AA{}  \\
$ l_h$ &  21.00 \AA{}   \\
 \multicolumn{4}{l}{Polaron shift} \\
$\Delta E_p^{e}$&  -45.3 meV \footnotemark[1]   \\
$\Delta E_p^{h}$ &  -49.2 meV \footnotemark[1]    \\
\end{tabular}
\end{ruledtabular}
\footnotetext[1]{Ref. \onlinecite{mapbi3_1}.}
\footnotetext[2]{Ref. \onlinecite{walsh2013}.}
\footnotetext[3]{Ref. \onlinecite{ramanmapbi3}.}
\end{table}

In the present work, the exciton energies are obtained solving the radial Schr\"odinger equation for the 
relative coordinate wave function of the exciton $\Phi(r)$
\begin{equation}
 \frac{{\rm d}^2\Phi}{{\rm d}r^2}+{\frac{2}{r}\frac{{\rm d}\Phi}{{\rm d}r}}+\left(\frac{2\mu}{\hbar^2}\left(E - V(r) \right){-\frac{l\left(l+1\right)}{r^2}}\right)\Phi=0, 
 \label{ec:radial}
\end{equation}
where $V(r)$ is the electron-hole interaction potential (Coulomb, Haken, or PB), $l$ is the azimuthal quantum 
number, of which we only consider $l=0$ that are the optically active states. 
The Eq. (\ref{ec:radial}) for $l=0$ 
has been solved 
integrating the equation starting from $r=0$ with the conditions $\Phi(0)>0$, $\Phi'(0)=0$ and imposing 
$\Phi(r_c)=0$, where $r_c$ is a cutoff radius sufficiently large to mimic the boundary conditions at infinity. 
The cutoff radii $r_c$ are established solving the equation for the Coulomb potentials and comparing the 
numerical energies with the known exact solutions. 
We have used exciton atomic units $a_0$ and $Ry_0$ for the radius and energy, respectively. 
The functions are normalized according to 
\begin{equation}
\int_0^{r_c} 4\pi |\Phi(r)|^2 r^2 dr = 1 .
\label{ec:norm}
\end{equation}

The optical oscillator strengths are defined as
\begin{equation}
f_n=\frac{2 m_0}{\hbar\omega_{n,0}}\vert \langle \Psi_n |\vec{\xi}\cdot\hat{\vec v} | 0 \rangle \vert^2 , 
\end{equation}
where $m_0$ is the free electron mass, $\hbar\omega_{n,0}=Eg^{*}+E_n$ is the transition energy, $|0\rangle$ and 
$|\Psi_n\rangle$ are the ground and excited states of the crystal, respectively, and 
$\hat{\vec v}=i [\hat{H},\vec{r} ]/\hbar$ is the velocity operator\cite{delsole93}. $E_g^{*}$ is the renormalized gap (with the polaron shift), and $E_n$ are the 
eigenvalues of Eq. (\ref{ec:radial}).
We shall approximate $f_n$ by the expression for pure excitons, i.e., neglecting the phonon coupling, as
\begin{equation}
f_n=\frac{2 m_0}{\hbar\omega_{n,0}}  \sum_{cv}\sum_{\alpha=x,y,z}   \frac{1}{3} \vert \langle u_{c\mathbf{0}} \vert \hat{v}_{\alpha} \vert u_{v\mathbf{0}} \rangle
\vert^2 \frac{\Omega_{f.u.}}{a_0^3} \vert\Phi_n(0)\vert^2  . 
\end{equation}
In the above expression, $\Omega_{f.u.}$ is the normalization volume of the center-of-mass part of the 
exciton envelope wave function function, which we consider as the volume of one formula unit, i.e.,  one fourth of the unit cell volume 
952.5 \AA$^3$. With this convention, the oscillator strength is equivalent to the values reported 
elsewhere\cite{ishihara94,Tanaka2003}. The factor $1/3$ and the sum in $\alpha$ correspond to isotropic average of the crystal orientations.
 $u_{v\mathbf{0}}$ and $u_{c\mathbf{0}}$ are the Bloch functions of the valence band maximum and conduction band minimum, 
which in this case are both doubly degenerate. Using first principles calculations (see the Appendix) we 
have calculated the parameter
\begin{equation}
U_{cv}=\frac{m_0}{2}\sum_{cv}\sum_{\alpha=x,y,z}   \frac{1}{3} \vert \langle u_{c\mathbf{0}} \vert \hat{v}_{\alpha} \vert u_{v\mathbf{0}} \rangle\vert^2  = 1.706 \mbox{ eV}.
\label{eq:ucv}
\end{equation}
Therefore, we obtain the simplified expression
\begin{equation}
f_n=\frac{4 U_{cv}}{\hbar\omega_{n,0}}   \frac{\Omega_{f.u.}}{a_0^3} \vert\Phi_n(0)\vert^2  . 
\label{ec:oscstr}
\end{equation}
Let us stress that the exciton Bohr radius $a_0^3$ appears in Eq.~(\ref{ec:oscstr}) 
only if the normalization condition (\ref{ec:norm}) is applied in relative units of $a_0$.


Both the Haken and Pollmann-B\"uttner potentials behave like a Coulomb potential for very large distance ($r\gg l_e,l_h$) or very short distances ($r\ll l_e,l_h$), screened by the low and high frequencies dielectric constants, respectively.  
Figure \ref{fig:plotpot} shows in logarithmic scale, the limiting Coulomb potentials screened by $\epsilon_0$ and $\epsilon_{\infty}$. These are represented by the straight lines, enclosing the 
Haken and Pollmann-B\"uttner potentials, that interpolate the limiting cases. Horizontal lines represent the eigenenergies of the exciton relative motion. 
The axes in the figure are in units of  static (fully screened) exciton radius and $a_0=\hbar^2\epsilon_0/\mu e^2$ and exciton energy $Ry_0=\mu e^4/2\hbar^2\epsilon^{2}$. In these units, the static Coulomb potential is given by $-2/r$ and the exciton eigenenergies are $E^0_{n}=-1/n^2$. The Coulomb 
potential and the hydrogenic energies defined by $\epsilon_{\infty}$ are $E^{\infty}_{n}=-\epsilon_r^2/ n^2$, where $\epsilon_r=\epsilon_{0}/\epsilon_{\infty}$. 
For the parameters  of  \mapbi{}  ($\epsilon_r=4.25$), one can appreciate in Figure \ref{fig:plotpot} and Table \ref{tab:en} that the lowest exciton levels are 
$E^{H}_n=-11.26$ and $E^{PB}_n=-8.65$ for the Haken and PB potentials.  These values represent a significant correction to  either $E_{1}^{0}=-1$ or $E^{\infty}_1=-18$. 
The excited exciton energies of Haken and PB potentials approach the values $-1/n^2$ for high $n$. 

In order to compare the energies $E^{H}_1$ and $E^{PB}_1$ one must consider that  $Ry_0$ is defined either by the polaron or the bare reduced mass in the first and second model, respectively. In absolute units, $E^{H}_1=-37$~meV and $E^{PB}_1=-24$~meV.
Is seems that the PB value is in better agreement with the experimental values near  
19~meV\cite{Sun2014perovskitas,portugall2015}.

\begin{table}
\caption{ \label{tab:en}
Exciton binding energies and wave function at origin. The wave function is normalized with the radii in units of $a_0$. In the same units,  the hydrogenic functions 
fulfill $\Phi^{0}_n(0)=1/\sqrt{\pi n^3}$.  In normal units, $\Phi_n(0)$ must be divided by $a_0^{3/2}$.}
\begin{ruledtabular}
\begin{tabular}{r|rr|rr}
  &  \multicolumn{2}{c|}{Haken model}  & \multicolumn{2}{c}{PB model}\\
$n$ &  $E^H_n/Ry_0$  &      $\Phi^H_n(0)$  &  $E^{PB}_n/Ry_0$  &      $\Phi^{PB}_n(0)$\\   
\hline
      1 & -11.2605 &   4.7255  &  -8.6473 &   4.5421 \\
       2 &  -0.8128 &   1.0832  &  -0.4622 &   0.6507 \\
       3 &  -0.2092 &   0.3636  &  -0.1661 &   0.3031 \\
       4 &  -0.0979 &   0.2047  &  -0.0839 &   0.1811 \\
       5 &  -0.0567 &   0.1358  &  -0.0505 &   0.1236 \\
       6 &  -0.0370 &   0.0985  &  -0.0337 &   0.0911 \\
       7 &  -0.0260 &   0.0756  &  -0.0240 &   0.0708 \\
       8 &  -0.0193 &   0.0604  &  -0.0180 &   0.0570 \\
       9 &  -0.0149 &   0.0497  &  -0.0140 &   0.0472 \\
      10 &  -0.0118 &   0.0418  &  -0.0112 &   0.0399 \\
      11 &  -0.0096 &   0.0358  &  -0.0092 &   0.0343 \\
      12 &  -0.0080 &   0.0311  &  -0.0076 &   0.0299 \\
      13 &  -0.0067 &   0.0274  &  -0.0065 &   0.0264 \\
      14 &  -0.0057 &   0.0243  &  -0.0055 &   0.0235 \\
      15 &  -0.0050 &   0.0218  &  -0.0048 &   0.0211\\
\end{tabular}
\end{ruledtabular}
\end{table}

\begin{figure}[htbp]
\begin{center}
\includegraphics[width=8.0cm]{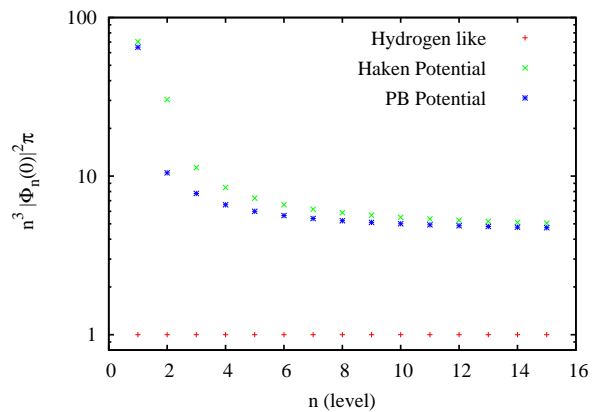}
\caption{Oscillator strengths of polaron excitons and hydrogenic exciton. \label{fig:os}}
\end{center}
\end{figure}

The values of $\Phi(0)$ in Table \ref{tab:en} and Figure \ref{fig:os} shows that the ground exciton 
is optically active. According to Eq.~(\ref{ec:oscstr}) the oscillator strength in the PB model is 0.013. 
The second exciton level is only $2.7$~meV (Haken) or $1.2$~meV (PB) below the edge of the continuum spectrum, and their oscillator strengths are one order of magnitude smaller than for the main line ($n=1$), making these transitions  practically undetectable in the optical spectra. Higher energies approach to the sequence $E_g-Ry_0/n^2$.
 For these higher levels, the oscillator strengths become proportional to the hydrogenic oscillator strengths.  The proportionality constant is fitted to $\beta_H=4.282$ (Haken) and $\beta_{PB}=4.214$ (PB) in each case. 
The fitting function was $f(n)=\beta+\gamma/(n-n_0)$, and the fitted $\beta$ are stable for any subset of
data with $n>7$. Let us note that 
\begin{equation}
\beta \simeq \frac{\epsilon_0}{\epsilon_{\infty}} . \label{eq:beta}
\end{equation}

This result, together with the approximation of the energies by the sequence $E_g^{*}-Ry_0/n^2$, leads to 
to a constant absorption spectrum near the band gap energy similar to 
the case of hydrogenic exciton\cite{pastori,Elliot}, 
\begin{equation}
\alpha(E_g^{*})=\frac{2\pi e^2 \hbar P_{cv}^2 }{E_g  n_r c m_0^2   } \frac{ \beta }{a_0^3 Ry_0 } ,
\end{equation}
where $n_r$ is the refraction index and can be approximated by $\sqrt{\epsilon_{\infty}}$. 
Evaluation the physical constants, and using Eqs.~(\ref{eq:beta}), (\ref{eq:ucv}), and (\ref{eq:pcv}), the above expression can be cast as 
\begin{equation}
\alpha(E_g^{*})=(3.466 \mbox{ nm}^{-1}) \frac{U_{cv}}{n_r E_g} \left( \frac{\mu}{m_0} \right)^2  \frac{ 1}{\epsilon_{\infty}} ,
\label{ec:absgap}
\end{equation}
Using the material parameters of \mapbi{}, $\alpha(E_g)\sim 3.1$~$\mu$m$^{-1}$.


The exciton spectrum obtained presents of a low energy non-hydrogenic state that is 37 or 24 meV (in 
Haken or PB models, respectively) below the onset of continuous free polaron spectrum.  The rest of the 
discrete exciton spectrum resembles the hydrogenic series with static dielectric constant $\epsilon_0$. The 
first exciton state dominates the absorption edge with oscillator strength $\sim 0.013$, while the oscillator 
strengths of higher states are one or two orders of magnitude smaller. Like the case of hydrogenic exciton, 
the coalescence of absorption lines from higher states generates a quasi-continuum spectrum with a finite 
value at the gap energy.  The absorption coefficient at gap energy in enhanced with respect to the 
hydrogenic model by a factor $\epsilon_0/\epsilon_{\infty}$. Extrapolating this behavior to the room temperature tetragonal 
phase, this enhancement factor could be responsible of the high capacity of this material to collect the photon energy in the photovoltaic cells. 

To interpret the absorption experiments and to determine de band gap, one needs to know 
wether the onset of the continuous absorption spectrum corresponds to the exciton 
continuum spectrum, or to the accumulation of discrete lines below the band gap. In other words, what is 
the energy range of constant absorption coefficient given by  Eq. (\ref{ec:absgap}).  Considering that the higher exciton levels are within 2 meV of the continuum, and  that the exciton absorption spectrum is dominated by the fundamental state, one can conclude that the band gap coincides with the absorption 
threshold after filtering the first exciton peak.    
On the other hand, the polaronic effect downshifts the gap by 95~meV. 
Therefore, the measured gap 1.68~eV should be understood as an electronic gap of 1.78~eV decreased 
by the polaron shift. 

Several kinds of estimations of the exciton binding energy in \mapbi{} have been reported. 
The first method was employed by Hirasawa et al\cite{hirasawa94a}, and repeated later with 
improved accuracy by Tanaka et al\cite{Tanaka2003}. They measured the exciton diamagnetic coefficient 
in magneto-absorption spectra, and related the measurements with the binding energy in the framework of 
the hydrogenic model with the high frequency dielectric constant. With this model, Tanaka et al determined a binding energy of 50 meV. The use of the high frequency dielectric constants was a choice of the model, and not determined by the experiments. Let us note that Tanaka et al, and Hirasawa et al used a value 
$\epsilon_{\infty}=6.5$ that is higher than our value. If our value $\epsilon_{\infty}=5.32$ were used, 
a binding energy of 65 meV would be obtained. Conversely, our binding energies using the larger dielectric constant would be smaller. These values of the exciton binding energy are strongly 
biased by the choice of the dielectric constants. 

The second method has been applied by Sun et al\cite{Sun2014perovskitas} ($E_B = 19$~meV). 
They  obtained the binding energy by fitting the photoluminescence intensity as a function of 
temperature with an Arrhenius equation, not using any model of the exciton states. It  is rewarding that  our calculation of the binding energy is in good agreement with the values determined by Sun et al. 
 Huang and Lambrecht\cite{lambrecht13} have argued, in a study of 
cesium tin halide perovskites, that the photoluminescence temperature dependence just 
give information on the free exciton linewidth or the binding energies of bound excitons, but not 
on free excitons. 
 However, Even et al\cite{evenjpcl14} fitted the absorption spectrum using the 
Wannier-Mott exciton model
and obtained a similar value for the binding energy, and reported an effective dielectric constant 
$\epsilon_{eff}=11$. This value of the dielectric constants, together with the assumed reduced 
mass $0.16 m_0$\cite{evenjpcl14} means a binding energy of 19 meV. 
Another method independent of the dielectric function has been used by Miyata et al\cite{portugall2015},  who performed magneto-absorption experiments with very high magnetic fields, determining a value of $16\pm 2$~eV. It is interesting that Miyata et al were able to detect the $2s$ exciton state for high magnetic field and extrapolated a $1s-2s$ difference of 15 meV at low magnetic field. Henceforth, assuming the hydrogenic model,  they estimated the binding energy in $20$ meV. However, extrapolating the Landau levels of the free exciton spectrum they obtained the precise value of $16$ meV. This observation agrees with our result that the $2s$ state is within $2$~meV of the free exciton edge.

The evaluation of the oscillator strength provides another argument against the model of 
hydrogenic Wannier excitons screened by $\epsilon_0$. Ishihara\cite{ishihara94}  reported an experimental values of 0.02, which is close to our value 0.013. If the ground exciton were well described by hidrogenic model with $\epsilon_0$, 
then $\Phi_{1}(0)=1/\sqrt{\pi}$ (compare with Table \ref{tab:en} and Fig.~\ref{fig:os}), and the oscillator strength would be $\sim 64$ times smaller. Therefore, at least for low temperature, the fundamental exciton state does 
not correspond to screening by $\epsilon_0$. 

We wish to stress that we have not fitted any parameter in this work, which would bring 
the exciton binding energies in closer agreement with the recent experimental results. The parameters 
with larger uncertainty are the dielectric constants and the LO phonon energy. 
The only available experimental value
of $\epsilon_{\infty}=6.5$\cite{hirasawa94a} is larger than the ab initio value 
used here and that value would reduce the calculated binding energy. The measurement 
of $\epsilon_{\infty}$ is rather old, with few published details, and a new determination
 for present-day thin films would be welcomed. The LO phonon energy $E_{LO}$ has been chosen from 
 the more prominent peak in the calculated Raman spectrum of \mapbi{}\cite{ramanmapbi3}, 
 which is close to LO phonon energies in II-VI and III-V semiconductors. As mentioned above, the model Hamiltonians were developed assuming a unique LO phonon energy. 
The Raman spectrum of \mapbi{} shows bands at lower wave numbers. Using and average energy  
of the Raman active peaks, which do not have necessarily LO character, may lead to lower exciton binding 
energy. An extension of the PB Hamiltonian to include several LO phonon branches 
would be a better founded approach.

In summary, we have calculated the exciton binding energies and oscillator strengths using two model
Hamiltonians of the exciton-phonon coupled system. The Pollmann-B\"uttner model Hamiltonian 
gives a binding energy in good agreement with recent experimental determinations. The calculated 
oscillator strength of the main exciton line agrees with the value estimated from experiments, while the strengths of higher transitions are much smaller.

\begin{acknowledgments}
We acknowledge computer time from the J\"ulich Supercomputing Centre (JSC) under the MOHP-SOPHIA project, and support from FONDECYT Grant. No. 1150538 and the European Project  NANOCIS  of the FP7-PEOPLE-2010-IRSES. We acknowledge J. C. Conesa, P. Palacios and C. Trallero-Giner for interesting discussions that motivated this work. 
\end{acknowledgments}

\appendix
\section{Interband matrix element}

With the VASP code\cite{vasp4} the tensor dielectric function is computed in the longitudinal approximation\cite{opticavasp}

\begin{eqnarray}
\varepsilon^{(2)}_{\alpha \beta} (\omega) & =&  \frac{4 \pi^2 e^2}{\Omega} \lim_{q\rightarrow 0} \frac{1}{q^2} 
\sum_{c,v,\mathbf{k}} g_S w_{\mathbf{k}}\delta(\epsilon_{c\mathbf{k+q}} - \epsilon_{v\mathbf{k}} - \hbar\omega )  \nonumber \\
& & \times \langle u_{c\mathbf{k}+\mathbf{e}_\alpha q} \vert u_{v\mathbf{k}} \rangle
\langle u_{c\mathbf{k}+\mathbf{e}_\beta q} \vert u_{v\mathbf{k}} \rangle^*,
\label{ec:longdiel}
\end{eqnarray}
where $w_k$ are the k-point weights, defined such that they sum to 1, $\epsilon_{c\mathbf{k+q}}$ are band energies, 
$\Omega$ is the unit cell volume, $m_0$ is the vacuum electron mass, $\mathbf{e}_\alpha$ are polarization vectors.  The factor $g_S$ is the spin degeneracy, which is 2 in Ref.~\onlinecite{opticavasp}, and the bands $v,c$ in the sum are restricted to have the same spin. In the calculation with spin-orbit coupling, we consider $g_S=1$ and the sum is over all pairs of valence and conduction bands.
The Eq.~\ref{ec:longdiel} is equivalent to the transverse approximation\cite{delsole93},
\begin{eqnarray}
\lefteqn{ 
\varepsilon^{(2)}_{\alpha \beta} (\omega)  =  \frac{4 \pi^2 e^2 }{\Omega\omega^2} 
\sum_{c,v,\mathbf{k}} g_S w_{\mathbf{k}}\delta(\epsilon_{c\mathbf{k}} - \epsilon_{v\mathbf{k}} - \hbar\omega )  }\nonumber \\
& \times  & \langle u_{c\mathbf{k}} \vert  \hat v_{\alpha} + \frac{\hbar k_{\alpha}}{m_0} \vert u_{v\mathbf{k}} \rangle
\langle u_{c\mathbf{k}} \vert \hat  v_{\beta} + \frac{\hbar k_{\beta}}{m_0} \vert u_{v\mathbf{k}} \rangle^*.
\end{eqnarray}

For a local Hamiltonians with spin-orbit coupling, $m_0\hat{\vec{v}}= -i \hbar\nabla + (\hbar/4m_0 c^2)\vec{\sigma} \times \nabla V$. The PAW potentials and the hybrid functionals introduce non-locality in the Hamiltonian, and the velocity operator contains additional terms\cite{delsole93}.  
For the purpose of the optical properties of the exciton, we only need the values of $\langle u_{c\mathbf{k}} \vert  \hat v_{\beta}  \vert u_{v\mathbf{k}} \rangle$ between the VBM and the CBM, which occurs at the $\Gamma$
 point ($k_{\beta}=0$). These values $\langle u_{c\mathbf{k}} \vert  \hat v_{\beta}  \vert u_{v\mathbf{k}} \rangle$ can be fitted from the dielectric function, which is calculated using (\ref{ec:longdiel}).  Hence, if the contribution of the $\Gamma$ point can be separated from the other k-point contributions, we have that
 
 \begin{eqnarray}
\varepsilon^{(2)}_{\alpha \beta;\Gamma} (\omega) & =&  \frac{4 \pi^2 e^2 }{\Omega\omega^2} 
 g_S w_{0}  \sum_{cv}^{\prime} \delta(\epsilon_{c\mathbf{0}} - \epsilon_{v\mathbf{0}} - \hbar\omega )  \nonumber \\
& & \times \langle u_{c\mathbf{0}} \vert \hat v_{\alpha} \vert u_{v\mathbf{0}} \rangle
\langle u_{c\mathbf{0}} \vert \hat v_{\beta}  \vert u_{v\mathbf{0}} \rangle^*.
\end{eqnarray}
 In the above expression, the sum $\sum^{\prime}$ is restricted to the top valence bands and bottom valence bands. 
 
 To fit with the exciton, we consider the averaged dielectric
  function $\varepsilon=(1/3)\mbox{Tr}\varepsilon_{\alpha\alpha}$
\begin{equation}
\varepsilon^{(2)}_{\Gamma} (\omega)  =  \frac{4 \pi^2 e^2 }{\Omega\omega^2 m_0^2} 
 g_S w_{0}  P_{cv}^2 \delta(\epsilon_{c\mathbf{0}} - \epsilon_{v\mathbf{0}} - \hbar\omega ), 
 \label{ec:fit}
\end{equation}
with 
\begin{equation}
P_{cv}^2 = \frac{1}{3}\sum_{\alpha} \sum_{cv}^{\prime}  \vert \langle u_{c\mathbf{0}} \vert m_0 \hat v_{\alpha} \vert u_{v\mathbf{0}} \rangle\vert^2.
\label{eq:pcv}
\end{equation}
The parameter $P_{cv}$ has dimension of momentum, and $P_{cv}^2/2m_0$ is the parameter $U_{cv}$ defined in Eq. (\ref{eq:ucv}).

In practical calculations, $\delta(\epsilon_{c\mathbf{0}} - \epsilon_{v\mathbf{0}} - \hbar\omega )$ is replaced 
by a smearing function. If gaussian smearing is used for the self-consistent calculation, the computed spectrum must be fitted with a Gaussian function weighted by  $4 \pi^2 e^2 
 g_S w_{0}  P_{cv}^2 /\Omega\omega^2 m_0^2 $. 
The ab initio calculation was performed sampling the Brillouin zone with a  $2\times 2\times 2$ k-point grid centered at the 
$\Gamma$ point, which is sufficient to obtain total energies and charge densities, but it is coarse for calculation of optical properties and it allows to isolate the contributions of the $\Gamma-$point transitions. 
With this k-point mesh, the  weight $w_{0}=1/8$. 
The details of the calculation are given in Ref.~\onlinecite{mapbi3_1}.


\end{document}